\documentclass[amsmath,12pt,amssymb,preprint,nofootinbib]{revtex4-2}
\usepackage{amsfonts} %\usepackage{citesort}
\usepackage{graphicx} % Include figure files
\usepackage{epsfig}
\usepackage{multirow}
\usepackage{bm}% bold math
\usepackage{color}
\usepackage{amssymb}
\usepackage{dsfont}
\usepackage{hyperref}
\usepackage{cleveref}
\newcommand{\diracslash}[1]{#1\llap{/\kern2pt}}

\newcommand{\be}{\begin{equation}}
	\newcommand{\ee}{\end{equation}}
\newcommand{\bea}{\begin{eqnarray}}
	\newcommand{\eea}{\end{eqnarray}}
\newcommand{\ba}[1]{\begin{array}{#1}}
	\newcommand{\ea}{\end{array}}

\newcommand{\bt}{\begin{tabular}}
	\newcommand{\et}{\end{tabular}}

\newcommand{\beas}{\begin{eqnarray*}}
	\newcommand{\eeas}{\end{eqnarray*}}

\begin{document}
\title{Impact of nonextensivity on the transport coefficients of strongly interacting QCD matter}
\author{Dhananjay Singh}
\email{snaks16aug@gmail.com}
\affiliation{Department of Physics, Dr. B R Ambedkar National Institute of Technology Jalandhar, 
	Jalandhar -- 144008, Punjab, India}

\author{Arvind Kumar}
\email{kumara@nitj.ac.in}
\affiliation{Department of Physics, Dr. B R Ambedkar National Institute of Technology Jalandhar, 
	Jalandhar -- 144008, Punjab, India}

	\begin{abstract}
		Tsallis nonextensive statistics is applied to study the transport coefficients of strongly interacting matter within the Polyakov chiral SU(3) quark mean field model (PCQMF). Nonextensivity is introduced within the PCQMF model through a dimensionless $q$ parameter to examine the viscous properties such as shear viscosity ($\eta$), bulk viscosity ($\zeta_b$), and conductive properties, including electrical conductivity ($\sigma_{el}$) and thermal conductivity ($\kappa$). Additionally, some key thermodynamic quantities relevant to the transport coefficients, like the speed of sound ($c_{sq}^2$) and specific heat at constant volume ($c_{vq}$), are calculated. The temperature dependence of the transport coefficients is explored through a kinetic theory approach with the relaxation time approximation. The results are compared to the extensive case where $q$ approaches 1. The nonextensive $q$ parameter is found to have a significant effect on all transport coefficients. We find that the nonextensive behaviour of the medium enhances both specific shear viscosity $\eta/s_q$ as well as conductive coefficients $\sigma_{el}/T$ and $\kappa/T^2$. In contrast, the normalised bulk viscosity $\zeta_b/s_q$ is found to decrease as the nonextensivity of the medium increases. We have also studied the transport coefficients for finite values of chemical potentials. The magnitude of $\eta$, $\sigma_{el}$, and $\kappa$ increases at lower temperatures while $\zeta$ is found to decrease for systems with non-zero chemical potential.

	\end{abstract}
	
	\maketitle
\newpage	
	\section{\label{intro}Introduction}
 
            Relativistic heavy-ion collision experiments offer a distinct opportunity to reproduce the extreme state of matter that existed at the universe's beginning, known as quark-gluon plasma (QGP). Experiments conducted at facilities like the Large Hadron Collider (LHC) at the European Organization for Nuclear Research (CERN) \cite{alice,cms,atlas} and the Relativistic Heavy Ion Collider (RHIC) at Brookhaven National Laboratory (BNL) \cite{brahms,phobos,phenix,star} have provided substantial evidence for the creation of QGP which has significantly advanced the exploration and understanding of the strongly interacting matter properties under extreme conditions of temperatures. Additionally, the Beam Energy Scan (BES) initiative at RHIC \cite{star2011,star2017}, along with current research programs at the Facility for Antiproton and Ion Research (FAIR) \cite{tahir,cbm} and the Nuclotron-based Ion Collider Facility (NICA) \cite{friman,keke}, are working to explore the characteristics of baryon-rich nuclear matter.  
            
            \par The role of transport coefficients is crucial for describing the evolution of the bulk matter created in relativistic heavy-ion collisions. These coefficients provide insight into how much a system deviates from ideal hydrodynamics and reveal important information about fluid dynamics and critical phenomena \cite{kss,csernai,romat,muronga,bhalerao}. Extracting these coefficients accurately from experimental data and evaluating them using various theoretical approaches is a prominent area of research today. The shear viscosity ($\eta$) of hot quantum chromodynamics (QCD) matter has garnered significant interest, primarily because of the surprisingly small value of the ratio of shear viscosity and entropy density, $\eta/s_q=1/4\pi$, which may resemble a nearly perfect fluid \cite{lacey2007}. This finding has led to speculation about the existence of a strongly interacting quark-gluon plasma (sQGP). However, the bulk viscosity to entropy density ratio, $\zeta_b/s_q$, has been suggested to increase near the critical temperature $T_c$ \cite{paech,kharzeev,karsch}. The decreasing value of $\eta/s_q$ and an increasing value $\zeta_b/s_q$ near $T_c$ is found to be consistent with the lattice calculations \cite{nakamura,meyer}. The electrical conductivity $\sigma_{el}$ is important for explaining the enhancement of low-mass dimuons \cite{mohanty2013} and serves as an essential input for magnetohydrodynamic simulations \cite{roy2015,roy2017}. Another key transport coefficient for hydrodynamic evolution at finite chemical potential is thermal conductivity $\kappa$, which has been explored in Refs. \cite{denicol2012,greif,denicol2014,kapusta}. These collective findings emphasise that transport coefficients play an important in measuring the properties of strongly interacting relativistic QCD matter and understanding its phase transitions \cite{csernai}.

            \par In recent years, transport coefficients have been extensively studied using a variety of effective QCD \cite{sasaki2010,marty,ghosh2015,ghosh2016,deb,tawfik} and hadronic models \cite{itakura,fernandez,lang,mitra,ghosh2014,ghosh20142,ghoshb,kadam}. 
            % Furthermore, other approaches, such as lattice QCD (LQCD) \cite{}, transport simulations \cite{}, and the functional renormalization group (FRG) \cite{} have made substantial contributions.  
            However, these models employ a statistical approach based on the Boltzmann-Gibbs (BG) statistics, which is valid only for systems having strong dynamical correlations, homogeneous and infinite heat bath, long-range interactions, and microscopic memory effects \cite{tsallis1988,tsallis1991,tsallis1997,tsallis2009,kodama}. However, in the initial stages of the heavy ion collision experiments, these conditions are rarely met. Hence, some quantities develop power-law-tailed distributions and become nonextensive. To address these issues, Tsallis proposed nonextensive statistics as a generalization of the BG statistics by introducing a dimensionless $q$ parameter to account for all potential variables that violate the assumptions of the standard BG statistics \cite{tsallis1988}. In this framework, Tsallis proposed a generalized non-additive entropy
            \begin{equation}
                s_q = \frac{1-\sum_{i=1}^{w}P_i^q}{q-1},
                \label{eq1}
            \end{equation}
            where $w$ is the number of microstates, $P_i$ is the probability distribution with $\sum_{i=1}^{w}P_i=1$, and $q$ is a positive real number called the nonextensive $q$-parameter. Assuming equiprobability ($P_i=1/w$), the Tsallis entropy in Eq. (\ref{eq1}) reduces to \cite{tsallis1988,gudima}
            \begin{equation}
                s_q = \frac{w^{1-q}-1}{1-q}=\ln_qw,
                \label{eq2}
            \end{equation}
            here the $q$-logarithm is defined as \cite{tsallis2014}
            \begin{equation}
                \ln_{q}(x)  \equiv \frac{x^{1-q}-1}{1-q},
                \label{eq3}
            \end{equation}
            with corresponding $q$-exponential expressed as 
            \begin{equation}
                \exp_{q}(x) = [1+(1-q)x]^{1/(1-q)}.
                \label{eq4}
            \end{equation}
            The non-additivity of the entropy $s_q$ follows from the non-additivity of the $q$-logarithm \cite{tsallis1988}: considering two independent systems $A$ and $B$ with $P_{(A+B)}=P_{A}P_{B}$, the generalized entropy of the system takes the form
            \begin{equation}
                s_q(A+B) = s_q(A)+s_q(B)+(1-q)s_q(A)s_q(B).
            \end{equation}
            The quantity $|1-q|$ quantifies the degree of the non-equilibration, i.e., how far the system is away from equilibrium. For $q>1$, it describes intrinsic fluctuations of temperature in the system \cite{wilk2000,biro2005}. In Ref. \cite{cleymans2009}, it was observed that temperature fluctuations are measured by the divergence of $q$ from unity, while the Boltzmann limit ($q$ = 1) does not show any temperature fluctuation. Thus, utilizing the nonextensive Tsallis statistics within the dynamical model is highly advantageous for examining the transport coefficients as they are not well-defined during the initial stage of heavy-ion collisions where the system is in non-equilibrium. In the present work, we attempt to see possible deviations from standard BG statistics for the values of $q>1$, as these values have been seen in numerous phenomenological investigations of high-energy heavy ion collisions \cite{cleymans2009,biyajima,cleymans2012}. In the limit $q\rightarrow1$, the nonextensive entropy reduces to the usual BG entropy, i.e., $s_{q=1}=s_{BG}$. The $q$-parameter is incorporated into the specific dynamical formulas of the model and enables a straightforward phenomenological test against possible deviations from the BG framework.

            Tsallis nonextensive statistics has gained considerable importance in recent years due to its ability to fit transverse momentum distributions across a broad range of collision energies, as demonstrated by the STAR \cite{abelev2007}, PHENIX \cite{phenix2011}, ALICE \cite{alice2011}, and CMS \cite{cms2011} collaborations. In light of this, nonextensivity has been incorporated into many theoretical models to study the properties of the QCD matter. These include a generalized Quantum Hydrodynamics ($q-$QHD) model \cite{pereira2007}, a nonextensive version of Nambu-Jona-Lasinio ($q-$NJL) model \cite{rozy2009}, a nonextensive version of MIT bag ($q-$MIT) model \cite{cardoso}, a generalized linear sigma model ($q-$LSM) \cite{shen}, and a nonextensive Polyakov chiral SU(3) quark mean field ($q-$PCQMF) model \cite{dj}. Recently, the $q-$PNJL model has been employed to study the transport coefficients \cite{zhao2020}, critical exponents \cite{zhao2021}, and fluctuations in the baryon number \cite{zhao2023}. Nonextensivity has also been incorporated within the relaxation time approximation of kinetic theory to study the viscous coefficients \cite{rath2024} and conductive coefficients \cite{rath2023} of hot and dense magnetized QCD matter. Furthermore, the bulk properties of protoneutron stars \cite{lavagno2011} and the thermodynamics of a black hole \cite{megias} have also been explored using nonextensive statistics. 

            \par In the present study, we aim to utilise the $q-$PCQMF   to investigate the transport coefficients of strongly interacting QCD matter at finite temperatures and chemical potentials. We study the temperature variations of the transport coefficients using the expressions obtained from kinetic theory and the relaxation time approximation at zero and finite values of chemical potentials. Additionally, we have included the presence of quark back reaction by replacing the usual Polyakov loop potential with the QCD glue potential \cite{lisa,suraj}. This is done by substituting the pure gauge temperature $T_{YM}$ with the glue potential temperature $T_{glue}$. This paper is organized as follows: In Sec. \ref{method}, we give a brief introduction of the nonextensive version of the Polyakov chiral SU (3) quark mean field ($q$-PCQMF) model. Sec. \ref{trans} discusses the expressions of the transport coefficients. The impact of the $q$-parameter on the transport coefficients of strongly interacting QCD matter is discussed in Sec. \ref{results}. Finally, our brief summary and conclusions are presented in Sec. \ref{summary}.
	\section{\label{method} $q$-PCQMF model}
            The thermodynamic potential density of the $q$-extended Polyakov chiral SU(3) quark mean field model in the mean field approximation is defined as \cite{dj}
            \begin{equation}
            \label{qtpd}
            \hspace*{-.4cm} 
            \Omega_{q}= \mathcal{U}(\Phi,\bar{\Phi},T) - {\cal L}_M- {\cal V}_{vac} + \sum_{i=u,d,s}\frac{-\gamma_i k_BT}{(2\pi)^3}\int_0^\infty 
            d^3k\left\{ {\rm ln_{q}} 
            F_{q}^{-}+
            {\rm ln_{q}} F^{+}_{q}\right\},
            \end{equation}
            where ${\cal L}_M = {\cal L}_{\Sigma\Sigma} +{\cal L}_{VV} +{\cal L}_{SB}$ is the meson interaction term. In this model, the attractive part of the interactions between quarks is represented by scalar meson fields $\sigma, \zeta,$ and $\delta$ while the repulsive part is represented by vector fields $\rho, \omega$ and $\phi$. The Polyakov loop fields $\Phi$ and $\bar{\Phi}$ are included in the model to study deconfinement phase transition. Additionally, the model incorporates broken scale invariance by introducing a scalar dilaton field, $\chi$ \cite{schechter,gomm,heide}. For the scalar meson, the self-interaction term ${\cal L}_{\Sigma\Sigma}$ is expressed in terms of the scalar fields as
            \begin{eqnarray}
            {\cal L}_{\Sigma\Sigma} =& -\frac{1}{2} \, k_0\chi^2
            \left(\sigma^2+\zeta^2+\delta^2\right)+k_1 \left(\sigma^2+\zeta^2+\delta^2\right)^2
            +k_2\left(\frac{\sigma^4}{2} +\frac{\delta^4}{2}+3\sigma^2\delta^2+\zeta^4\right)\nonumber \\ 
            &+k_3\chi\left(\sigma^2-\delta^2\right)\zeta 
            -k_4\chi^4-\frac14\chi^4 {\rm ln}\frac{\chi^4}{\chi_0^4} +
            \frac{d}
            3\chi^4 {\rm ln}\left(\left(\frac{\left(\sigma^2-\delta^2\right)\zeta}{\sigma_0^2\zeta_0}\right)\left(\frac{\chi^3}{\chi_0^3}\right)\right),
            \label{scalar0}
            \end{eqnarray}
            where $\sigma_0 = - f_\pi$ and $\zeta_0  = \frac{1}{\sqrt{2}} ( f_\pi - 2 f_K)$ correspond to the vacuum values of $\sigma$ and $\zeta$ field, where $f_{\pi} =$ 93 MeV and $f_K =$ 115 MeV are the pion and kaon decay constant, respectively. The value of $d = 6/33$  is chosen to produce the correct trace anomaly for three flavours and three colours of quarks \cite{wang2003}. The vector meson self-interaction term is given by
            \begin{equation}
            {\cal L}_{VV}=\frac{1}{2} \, \frac{\chi^2}{\chi_0^2} \left(            m_\omega^2\omega^2+m_\rho^2\rho^2+m_\phi^2\phi^2\right)+g_4\left(\omega^4+6\omega^2\rho^2+\rho^4+2\phi^4\right), \label{vector}
            \end{equation}
            with $m_{\phi}=$ 1020 MeV is the $\phi$ meson mass and $m_{\omega}=m_{\rho}=$ 783 MeV is the mass of $\omega$ and $\rho$ meson. Finally, the spontaneous symmetry-breaking term ${\cal L}_{SB}$ is written as \cite{wang2002}
            \begin{equation} 
                {\cal L}_{SB}=-\frac{\chi^2}{\chi_0^2}\left[m_\pi^2f_\pi\sigma + 
                \left(\sqrt{2}m_K^2f_K-\frac{m_\pi^2}{\sqrt{2}}f_\pi\right)\zeta\right].
                \label{esb_ldensity}
            \end{equation} The term $\mathcal{U}(\Phi,\bar{\Phi}, T)$ in Eq. (\ref{qtpd}) is the Polyakov loop effective potential, which in the logarithmic form is given by \cite{fukushima,fukugita}
            \begin{eqnarray}
                \frac{{\cal U}(\Phi,\bar{\Phi})}{T^4}&=&-\frac{a(T)}{2}\bar{\Phi}\Phi+b(T)\mathrm{ln}\big[1-6\bar{\Phi}\Phi+4(\bar{\Phi}^3+\Phi^3)-3(\bar{\Phi}\Phi)^2\big],
    	\label{log}
            \end{eqnarray}
            where
            \begin{equation}\label{T}
	           a(T)=a_0+a_1\bigg(\frac{T_0}{T}\bigg)+a_2\bigg(\frac{T_0}{T}\bigg)^2,\ \            b(T)=b_3\bigg(\frac{T_0}{T}\bigg)^3,
            \end{equation}
            with $a_0 = 1.81$, $a_1 = -2.47$, $a_2 = 15.2$ and $b_3 = -1.75$ \cite{manisha}. Incorporating the effects of the backreaction of quarks leads to substituting the Polyakov loop potential with the QCD glue potential \cite{lisa}. Denoting the Polyakov loop potential in Eq. (\ref{log}) as ${\cal U}_{YM}$, the improved glue Polyakov loop potential ${\cal U}_{glue}$ is written as \cite{suraj} 
            \begin{equation}\label{Uglue}
	           \frac{{\cal U}_{glue}(\Phi,\bar{\Phi},T_{glue})}{T_{glue}} = \frac{{\cal U}_{YM}(\Phi,\bar{\Phi},T_{YM})}{T_{YM}},
            \end{equation}
            with $T = T_{0}^{YM}\left(1 + 0.57\left(\frac{T_{glue}}{T_{0}^{glue}-1}\right)\right)$ and $T_0 = T_{0}^{YM}$ in the RHS of the Eq. (\ref{log}). In the present work, we have taken $T_{0}^{YM} = T_{0}^{glue} = 200$ MeV. In the last term of Eq. (\ref{qtpd}), $\gamma_i = 2$ is the spin degeneracy factor, and
            \begin{eqnarray}
                F_{q}^{-}=&1+\exp_{q}(-3E^-)+3\Phi \exp_{q}(-E^-)+3\bar{\Phi}\exp_{q}(-2E^-), \\
                F_{q}^{+}=&1+\exp_{q}(-3E^+)+3\bar{\Phi} \exp_{q}(-E^+)+3\Phi \exp_{q}(-2E^+),
            \end{eqnarray}
            where $E^- = (E_i^*(k)-{\mu_i}^{*})/k_BT$ and $E^+ = (E_i^*(k)+{\mu_i}^{*})/k_BT$ and $E_i^*(k)=\sqrt{m_i^{*2}+k^2}$ represents the effective energy of quarks. The in-medium quark chemical potential $\mu_i^*$ can be written in terms of quark chemical potential $\mu_i$ as ${\mu_i}^{*}=\mu_i-g_\omega^i\omega-g_\phi^i\phi-g_\rho^i\rho$. Here, $g^i_{\omega}$, $g^i_{\phi}$ and $g^i_{\rho}$ are the coupling coefficients between vector meson fields and various quarks. The in-medium mass of quarks ${m_i}^{*} = -g_{\sigma}^i\sigma - g_{\zeta}^i\zeta - g_{\delta}^i\delta + \Delta m_i$, with $g_{\sigma}^i$, $g_{\zeta}^i$, and $g_{\delta}^i$ represent the coupling constants between scalar meson fields and various quarks and $\Delta m_{u,d} = 0$ 
            and $\Delta m_s = 29$ MeV. The term ${\cal V}_{vac}$ in Eq. (\ref{qtpd}) is subtracted to get zero vacuum energy. The temperature dependence of the scalar and vector fields is obtained by minimizing the thermodynamic potential density in Eq. (\ref{qtpd}) with respect to these fields, i.e.,
            \begin{equation}
            \label{minimize}
            \frac{\partial\Omega_q}
            {\partial\sigma}=\frac{\partial\Omega_q}{\partial\zeta}=\frac{\partial\Omega_q}{\partial\delta}=\frac{\partial\Omega_q}{\partial\chi}=\frac{\partial\Omega_q}{\partial\omega}=\frac{\partial\Omega_q}{\partial\rho}=\frac{\partial\Omega_q}{\partial\phi}=\frac{\partial\Omega_q}{\partial\Phi}=\frac{\partial\Omega_q}{\partial\bar\Phi}=0.
            \end{equation}
            The resulting coupled equations are provided in the Appendix. The vector density $\rho_{q,i}$ and the scalar density $\rho_{q,i}^s$ in the $q-$PCQMF model are defined as
            \begin{equation}
                \rho_{q,i} = \gamma_{i}N_c\int\frac{d^{3}k}{(2\pi)^{3}}  
                \Big(f_{q,i}(k)-\bar{f}_{q,i}(k)
                \Big),
            \label{rhov0}
            \end{equation}
            and
            \begin{equation}
                \rho_{q,i}^{s} = \gamma_{i}N_c\int\frac{d^{3}k}{(2\pi)^{3}} 
                \frac{m_{i}^{*}}{E^{\ast}_i(k)} \Big(f_{q,i}(k)+\bar{f}_{q,i}(k)
                \Big),
            \label{rhos0}
            \end{equation}
            respectively, with $q$-modified Fermi-distribution functions for quarks and antiquarks
            \begin{equation}\label{qdistribution}
            f_{q,i}(k)=\frac{\Phi \exp^{q}_{q}(-E^-)+2\bar{\Phi} \exp^{q}_{q}(-2E^-)+\exp^{q}_{q}(-3E^-)}
            {[1+3\Phi \exp_{q}(-E^-)+3\bar{\Phi} \exp_{q}(-2E^-)+\exp_{q}(-3E^-)]^{q}} , 
            \end{equation}            
            \begin{equation}\label{qdistribution1}
            \bar{f}_{q,i}(k)=\frac{\bar{\Phi} \exp^{q}_{q}(-E^+)+2\Phi \exp^{q}_{q}(-2E^+)+\exp^{q}_{q}(-3E^+)}
            {[1+3\bar{\Phi} \exp_{q}(-E^+)+3\Phi \exp_{q}(-2E^+)+\exp_{q}(-3E^+)]^{q}},
            \end{equation}
            respectively. 

            It is crucial to mention that as \( q \) approaches 1, the standard Fermi-distribution functions are restored, leading us back to the conventional (extensive) PCQMF model. Additionally, as temperature, $T\rightarrow 0$, the $q$-dependent thermodynamic potential density $\Omega_q$ defined in Eq. (\ref{qtpd}) also returns to its standard (extensive) form defined in Ref. \cite{manisha}, as long as $q>1$. This implies that the nonextensive effects are more prominent in heavy-ion collision experiments where the temperature reaches a few orders of MeVs and the value of $q$ remains greater than 1 \cite{li2013,cleymans2013,azmi}.

            The parameters of the model used in the present study are summarized in Table \ref{tab1}. These are adjusted to accurately reproduce the vacuum masses of $\pi$, $K$, $\sigma$, $\zeta$, $\chi$, and the average masses of $\eta$ and $\eta^{'}$ \cite{wang2003}. The relations of the baryon, isospin, and strangeness chemical potential are defined as
            \begin{eqnarray}
            \mu_B &=& \frac{3}{2}(\mu_u+\mu_d), \\
            \mu_I &=& \frac{1}{2}(\mu_u-\mu_d), \\
            \mu_S &=& \frac{1}{2}(\mu_u+\mu_d-2\mu_s),
            \end{eqnarray}
            respectively. Here, $\mu_u$, $\mu_d$, and $\mu_s$ represent the chemical potentials of up, down, and strange quarks, respectively.
            \begin{table}
            \centering
            \begin{tabular}{|c|c|c|c|c|c|c|c|c|c|}
            \hline
            $k_0$           & $k_1$          & $k_2$          & $k_3$         & $k_4$         & $g_s$         & $\rm{g_v}$          & $\rm{g_4}$           & $d$          & $\rho_0$(fm$^{-3}$)                            \\ \hline
            4.94                 & 2.12                & -10.16              & -5.38              & -0.06              & 4.76               & 6               & 37.5                 & 0.18               & 0.15                                  \\ \hline
            $\sigma_0$ (MeV) & $\zeta_0$(MeV)  & $\chi_0$(MeV)   & $m_\pi$(MeV)  & $f_\pi$(MeV)  & $m_K$(MeV)    & $f_K$(MeV)     & $m_\omega$(MeV) & $m_\phi$(MeV)  & $m_\rho$( MeV)                   \\ \hline
            -93                  & -96.87              & 254.6               & 139                & 93                 & 496                & 115                 & 783                  & 1020                & 783                                   \\ \hline
            $g_{\sigma}^u$  & $g_{\sigma}^d$ & $g_{\sigma}^s$ & $g_{\zeta}^u$ & $g_{\zeta}^d$ & $g_{\zeta}^s$ & $g_{\delta}^u$ & $g_{\delta}^d$  & $g_{\delta}^s$ & $g^u_{\omega}$ \\ \hline
            3.36                 & 3.36                & 0                   & 0                  & 0                  & 4.76               & 3.36                & -3.36                & 0                   &      3.86                             \\ \hline
            $g^d_{\omega}$ & $g^s_{\omega}$ & $g^u_{\phi}$ &  $g^d_{\phi}$ & $g^s_{\phi}$  & $g^u_{\rho}$   & $g^d_{\rho}$  &   $g^s_{\rho}$  &    &                               \\ \hline
            3.86       &        0         &           0         &    0               &           5.46        &      3.86          &    -3.86             &          0       &                    &                                   \\ \hline
            \end{tabular}
            \caption{The list of parameters used in the present work \cite{wang2003}.}
            \label{tab1}
            \end{table}

            \section{\label{trans} Transport coefficients}
            Transport coefficients for a system in the hydrodynamic regime can be determined using the Kubo formalism \cite{kubo,ghosh2019}, assuming that the relaxation time is shorter than the system's lifetime.
  The expressions for the transport coefficients obtained using this formalism are identical to those derived within a quasiparticle approach in kinetic theory using the relaxation time approximation (RTA) \cite{arnold,chakraborty}. In kinetic theory, the transport coefficients are derived using the Boltzmann transport equation, which can be written in the relaxation time approximation (RTA) as
            	\begin{equation}
            		k^{\mu}\partial_{\mu}f_{i}=C[f],
            	\end{equation}
            	where $C[f]$ is the collision integral. To study the transport coefficients, we are interested in small departures of the distribution function from the equilibrium,
            	\begin{equation}
            		\delta f_{i}(\vec x,\vec k,t) = f_{i}^{'}(\vec x,\vec k,t) - f_{i}(\vec x,\vec k,t).
            	\end{equation}
            	Here $f_i$ is the local equilibrium distribution of quarks, and $f^{'}_{i}$ is the non-equilibrium distribution function. Under nonextensive statistics, the equilibrium distributions are modified to their $q-$modified versions. This results in $q-$generalized transport equation known as nonextensive Boltzmann transport equation (NEBE) \cite{lavagno2002},
            	\begin{equation}
            		k^{\mu}\partial_{\mu}f_{q,i}=C_q[f],
            	\end{equation}
            	with $C_q[f]$ being the $q-$deformed collision term. 
            	%It is important to note that thermodynamic consistency requires that both the collision term $C_q[f]$ as well as $k^{\mu}\partial_{\mu}f^q$ must be explicitly modified \cite{lavagno2002}.
            	 In Ref. \cite{biro2012}, authors demonstrated that it is valid to employ conventional methods for calculating transport coefficients, beginning with NEBE. These computations yield relations for all transport coefficients that are formally analogous to those derived from the conventional Boltzmann-Gibbs distributions.          
%            
%            The expressions for the transport coefficients obtained using this formalism are identical to those derived within a quasiparticle approach in kinetic theory using the relaxation time approximation (RTA) \cite{arnold,chakraborty}. 
The expressions of various transport coefficients used in the present work are presented below \cite{ghosh2019,saha2018,islam}:
            \begin{eqnarray}
                \eta&=&\frac{2N_{c}}{15T}\sum_{i=u,d,s}\int\frac{{\rm d}^3k}{(2\pi)^3}\tau\left(\frac{k^{2}}{E_i^{*}}\right)^{2}[f_{q,i}(1-f_{q,i})+\bar{f}_{q,i}(1-\bar{f}_{q,i})],
                \label{eta}
            \end{eqnarray}
            \begin{eqnarray}
                \zeta_b&=&\frac{2N_{c}}{T}\sum_{i=u,d,s}\int\frac{{\rm d}^3k}{(2\pi)^3}\tau\frac{1}{E_i^{*2}}\left[\left(\frac{1}{3}-c_{sq}^{2}\right)k^{2}-c_{sq}^{2}m_i^{*2}+c_{sq}^{2}m_i^{*}T\frac{dm_i^{*}}{dT}\right]^{2} \nonumber \\ 
                &&\left[f_{q,i}(1-f_{q,i})+\bar{f}_{q,i}(1-\bar{f}_{q,i})\right], 
            \label{zeta}
            \end{eqnarray}
            \begin{eqnarray}
                \sigma_{el}&=&\frac{2N_{c}}{3T}\sum_{i=u,d,s}e_i^{2}\int\frac{{\rm d}^3k}{(2\pi)^3}\tau\left(\frac{k}{E_i^{*}}\right)^{2}[f_{q,i}(1-f_{q,i})+\bar{f}_{q,i}(1-\bar{f}_{q,i})],
            \label{sigma}
            \end{eqnarray}
            \begin{eqnarray}
                \kappa &=& \frac{2N_c}{3T^2}\sum_{i=u,d,s}\int\frac{{\rm d}^3k}{(2\pi)^3}\tau\left(\frac{k}{E_i^*}\right)^2[(E_i^*-h_q)^2f_{q,i}(1-f_{q,i})+(E_i^*+h)^2 \nonumber \\
                &&\bar{f}_{q,i}(1-\bar{f}_{q,i})].
            \label{kappa}
            \end{eqnarray}
            
            Notably, $f_{q, i}$ and $\bar{f}_{q, i}$ are not the standard Fermi-distribution functions but rather the $q$-version of the Fermi-distribution functions, given by Eqs. (\ref{qdistribution}) and (\ref{qdistribution1}). $c^2_{sq}$ is the speed of sound at constant entropy defined as 
            % $c_{sq}^{2} = \left(\frac{\partial p_q}{\partial \epsilon_q}\right)_{s_{q}} = \frac{s_{q}}{c_{vq}}$
            $c_{sq}^{2} = \left(\partial p_q/\partial \epsilon_q\right)_{s_{q}} = s_{q}/c_{vq}$ and $c_{vq}$ is the specific heat at constant volume defined as 
            % $c_{vq} = \left(\frac{\partial \epsilon_q}{\partial T}\right)_V$
            $c_{vq} = \left(\partial\epsilon_q/\partial T\right)_V$. The pressure is given by $p_q = -\Omega_q$ while the energy density and the entropy density are defined as $\epsilon_q=\Omega_q+\sum_{i=u,d,s} \mu_i^* \rho_i+Ts_q$ and 
            % $s_q=-\frac{\partial \Omega_q}{\partial T}$
            $s_q=-\partial \Omega_q/\partial T$, respectively. The heat function $h_q = (\epsilon_q + p_q)/\rho_q$ diverges at $\mu=0$ where $\rho_q$ diverges. The relaxation time $\tau$ is the measure of the timescale over which the distribution function relaxes back to equilibrium and is defined as \cite{hosoya} 
            \begin{equation}
                \tau = \frac{1}{5.1T\alpha_S^2\log(\frac{1}{\alpha_S})(1+0.12(2N_f+1))},
            \label{tau}
            \end{equation}
            where $\alpha_s$ is the temperature and chemical potential dependent strong coupling constant given by \cite{bannur,zhu}
            \begin{equation}
                \alpha_S(T,\mu)=\frac{6\pi}{(33-2N_f)\log\left(\frac{T}{\Lambda_T}\sqrt{1+(\frac{\mu}{\pi T})^2}\right)}\left(1-\frac{3(153-19N_f)}{(33-2N_f)^2}\frac{\log\left(2\log\frac{T}{\Lambda_T}\sqrt{1+(\frac{\mu}{\pi T})^2}\right)}{\log\left(\frac{T}{\Lambda_T}\sqrt{1+(\frac{\mu}{\pi T})^2}\right)}\right),
            \label{alpha}
            \end{equation}
            with $\Lambda_T=70$ MeV \cite{zhu}. 
      
	\section{\label{results}Results and Discussions}
        In this section, we will discuss the impact of the nonextensive $q$ parameter on the thermodynamic quantities and transport coefficients of strongly interacting QCD matter within the $q-$PCQMF model presented in Sec. \ref{method}. The $q$-PCQMF model incorporates the influence of the $q$ parameter through the thermodynamic potential density, $\Omega_q$. This modification of the potential density results in modification of the scalar density, $\rho_{q,i}^{s}$, and vector density, $\rho_{q,i}$, of constituent quarks. These densities, in turn, modifies the scalar ($\sigma$, $\zeta$, $\delta$, $\chi$), vector ($\omega$, $\rho$, $\phi$), and Polyakov loop fields ($\Phi$, $\bar{\Phi}$) which are determined by solving the interconnected system of non-linear equations given in Appendix.
        \par
\begin{figure}
                \centering
                \includegraphics[]{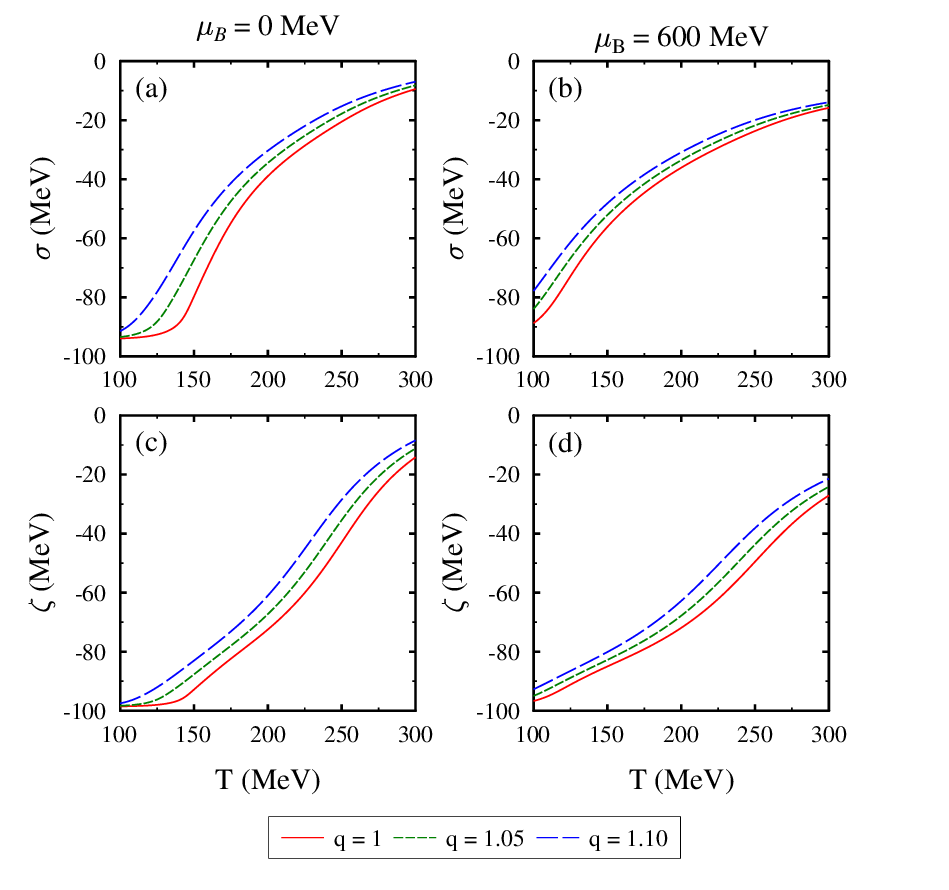}
                \caption{The scalar fields $\sigma$ and $\zeta$ are plotted as a function of temperature $T$ for the value of nonextensivity parameter $q$ = 1, 1.05, and 1.10, at baryon chemical potential $\mu_B = 0$ MeV [in subplots (a) and (c)], and baryon chemical potential $\mu_B = 600$ MeV, isospin chemical potential $\mu_I = -30$ MeV, and strangeness chemical potential $\mu_S = 125$ MeV [in subplots (b) and (d)].}
                \label{fields}
            \end{figure}

 Let us start by examining how the nonextensive $q$ parameter impacts the scalar fields $\sigma$ and $\zeta$. Figure \ref{fields} displays the changes in the scalar fields $\sigma$ and $\zeta$, with respect to temperature $T$, while keeping the baryon chemical potentials $\mu_B$ fixed at 0 and 600 MeV, respectively. The values of $q$ used are 1, 1.05, and 1.10. It is evident that the magnitude of $\sigma$ and $\zeta$ fall as the temperature of the medium rises. The decrease in the amplitude of scalar fields may indicate the restoration of chiral symmetry at elevated temperatures. As the temperature rises, the Fermi distribution function described by Eqs. (\ref{qdistribution}) and (\ref{qdistribution1}) decreases, leading to a reduction in the magnitude of the scalar fields. As the value of $q$ becomes greater than 1, at a certain temperature $T$, the scalar fields $\sigma$ and $\zeta$ experience a decrease in magnitude. For example, at zero baryon chemical potential and temperature $T=200$ MeV, the magnitude of the $\sigma$ ($\zeta$) field decreases from 38.97 MeV (72.49 MeV) at $q=1$ to 34.49 MeV (67.27 MeV) and 30.34 MeV (61 MeV) as the value of $q=$ 1.05 and 1.10, respectively. This indicates that the restoration of chiral symmetry at higher temperatures is faster for systems with a higher degree of nonextensivity, i.e., higher $q$.
        % The pseudo-critical temperature, denoted as $T_p$, is the temperature at which the magnitude of scalar fields starts to fall rapidly. Here, it is evident that as the value of $q$ increases, the value of $T_p$ decreases. For larger values of $q$, there is a significant reduction in the strength of the fields, even at lower temperatures. This suggests that the transition temperature has been adjusted to a lower value.
        Additionally, in subplots (b) and (d) of Fig. \ref{fields}, temperature variations of the scalar fields $\sigma$ and $\zeta$ are shown at a finite value of the baryon chemical potential $\mu_B = 600$ MeV, isospin chemical potential $\mu_I = -30$ MeV, and strangeness chemical potential $\mu_S=125$ MeV, for $q$ = 1, 1.05, and 1.10. At low temperatures, when the baryon chemical potential is raised from zero to a non-zero value, we notice a drop in the magnitude of the scalar fields. At temperature $T=120$ MeV and $q=1$, as the baryon chemical potential is increased from 0 to 600 MeV, the magnitude of the $\sigma$ ($\zeta$) field decreases from 93.1 MeV (98.2 MeV) to 76.73 MeV (92.51 MeV), respectively. Again, increasing the degree of nonextensivity, i.e., the magnitudes of $\sigma(\zeta)$ decrease to 70.22 MeV (90.11 MeV) and 64.46 MeV (87.75 MeV) for $q=1.05$ and 1.10, respectively. Ultimately, we can deduce that the chiral symmetry restoration shifts to lower temperatures as the value of baryon chemical potential becomes finite.        
        \par The order parameters used to investigate deconfinement in the mean-field approximation are the Polyakov loop fields, denoted as $\Phi$ and $\bar{\Phi}$. Temperature variations in the magnitude of $\Phi$ and $\bar{\Phi}$ provide insights into the deconfinement phase transition. Figure \ref{phi} shows the Polyakov fields $\Phi$ and $\bar{\Phi}$ as a function of temperature $T$, with baryon chemical potential set at $\mu_B=0$ and 600 MeV, for $q=1$, 1.05. and 1.10. At zero baryon chemical potential, in subplots (a) and (c) of Fig. \ref{phi}, we observe that the values of $\Phi$ and $\bar{\Phi}$ are almost zero at lower temperatures, suggesting the presence of confined quarks within hadrons. As the temperature increases, quarks transition from being confined to becoming deconfined, and the values of $\Phi$ and $\bar{\Phi}$ become nonzero. Like scalar fields, we observe that as nonextensivity grows (with greater values of $q$), the increase in $\Phi$ and $\bar{\Phi}$ happens earlier. This suggests a reduction in the temperature at which deconfinement occurs for values of $q$ greater than 1. For a finite baryon chemical potential of $\mu_B=600$ MeV (Figs. \ref{phi}(b) and \ref{phi}(d)), the Polyakov fields, $\Phi$ and $\bar{\Phi}$, have non-zero values even at lower temperatures. This may imply that quarks become deconfined at lower temperatures for non-zero values of baryon chemical potential. In conclusion, increasing nonextensivity or baryon chemical potential results in shifting the deconfinement temperature to lower values.
        \par
\begin{figure}
                \centering
                \includegraphics[scale=0.85]{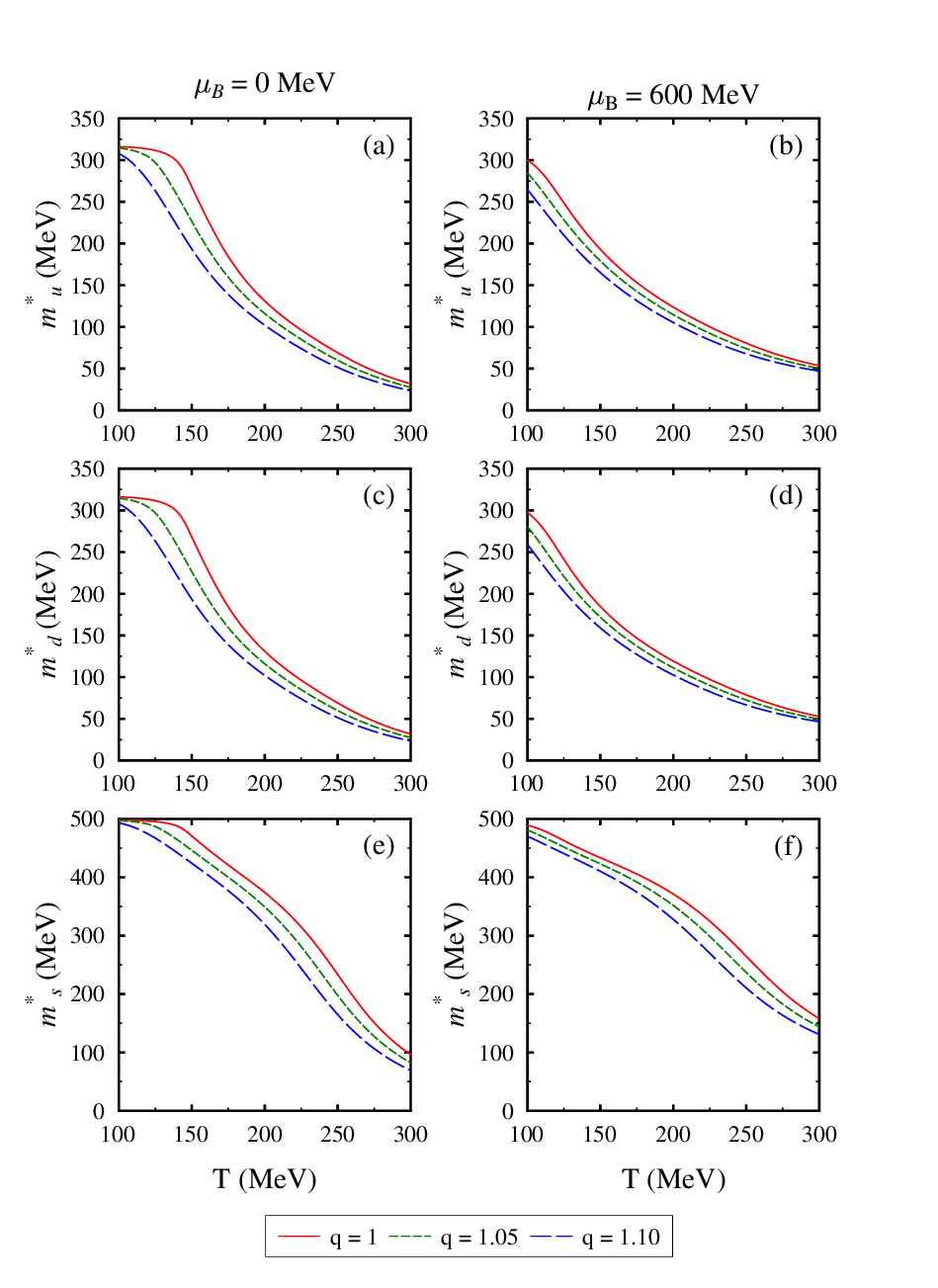}
                \caption{The effective quark masses $m_u^*,m_d^*,$ and $m_s^*$ plotted as a function of temperature $T$ for the value of nonextensivity parameter $q$ = 1, 1.05, and 1.10, at baryon chemical potential $\mu_B = 0$ MeV [in subplots (a), (c), and (e)], and baryon chemical potential $\mu_B = 600$ MeV, isospin chemical potential $\mu_I = -30$ MeV, and strangeness chemical potential $\mu_S = 125$ MeV [in subplots (b),(d), and (f)].}
                \label{mass}
            \end{figure}

 Next, we discuss the impact of the $q$ parameter on the in-medium masses of quarks. Figure \ref{mass} shows the temperature variations of effective masses of quarks $m_u^*, m_d^*,$ and $m_s^*$ at baryon chemical potentials $\mu_B = 0$ and 600 MeV, for $q=1$, 1.05. and 1.10. The in-medium effective quark masses are dependent on the in-medium scalar fields. As discussed in Fig. \ref{fields}, the magnitude of the scalar fields $\sigma$ and $\zeta$ decreases as the temperature of the medium is raised. This results in the fall of the effective quark masses with a rise in temperature. This may be attributed to the transition of quarks from within the confined state of hadrons to a state of deconfined QGP at higher temperatures. For a given temperature, we observe that increasing the nonextensivity ($q>1$) results in a decrease in the effective quark masses. Again, this may be indicative that the chiral symmetry restoration is shifted to lower temperatures with increasing nonextensivity in the system, as was discussed in Fig. \ref{fields}. As for the case of finite baryon chemical potential of $\mu_B = 600$ MeV (subplots (b), (d), and (f) of Fig. \ref{mass}), we find that the quark masses are reduced even at lower temperature values. Also, due to a finite value of isospin chemical potential $\mu_I = -30$ MeV, there is a minor difference in the effective masses of $u$ and $d$ quarks. For example, at $T=200$ MeV and $q=1$, we find the effective mass of $u$ quark to be $m^*_u \approx 123$ MeV while $m^*_d\approx119$ MeV. For values of $q$ greater than 1, we observe these values to decrease more sharply, indicating a quicker restoration of chiral symmetry for systems with a higher degree of nonextensivity. 
        \par
\begin{figure}
                \centering
                \includegraphics[scale=0.68]{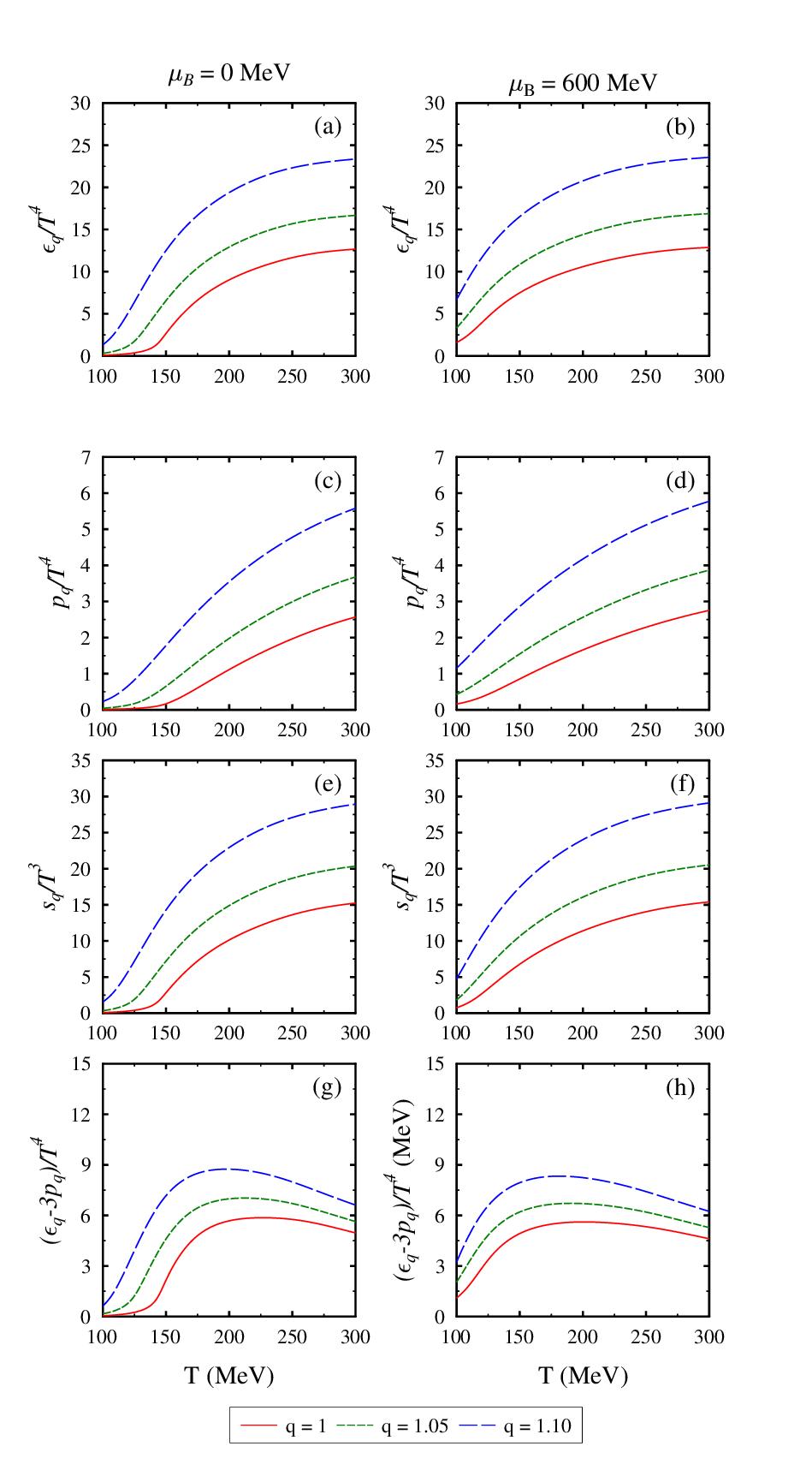}
                \caption{The scaled pressure density $p_q/T^4$, scaled energy density $\epsilon_q/T^4$, scaled entropy density $s_q/T^3$, and trace anomaly $(\epsilon_q-3p_q)/T^4$ as a function of temperature $T$ for the value of nonextensivity parameter $q$ = 1, 1.05, and 1.10, at $\mu_B = 0$ MeV [in subplots (a), (c), (e), and (g)] and $\mu_B=600$ MeV, $\mu_I = -30$ MeV, $\mu_S=125$ MeV [in subplots (b), (d), (f), and (h)].}
                \label{thermo}
            \end{figure}

 Figure \ref{thermo} showcase the changes in scaled thermodynamic quantities: $\epsilon_q/T^4, p_q/T^4,s_q/T^3,$ and $(\epsilon_q-3p_q)/T^4$, as temperature $T$ varies. The results are shown for both zero and as well as finite value of baryon chemical potential at $q = 1, 1.05,$ and 1.10. All of these quantities remain insignificant at low temperatures, but they increase when the system approaches the transition point. At high temperatures, the interaction among quarks weakens, causing the thermodynamic properties to approach the ideal gas limit or the Stefann-Boltzmann (SB) limit \cite{borsanyi,manisha}. Notably, for $q=1$, all of them stay under their respective SB limit at high temperatures. However, when $q>1$, we find that these thermodynamic quantities grow rapidly and go beyond their respective SB limits at higher temperatures. A similar impact of nonextensivity was observed on the thermodynamic quantities in the PNJL model \cite{zhao2020}. This is a consequence of using Tsallis statistics in the model. As when $T \rightarrow \infty$, the $q-$modified potential density $\Omega_q$ does not converge to its standard value $\Omega$. This means that for systems with $q>1$, the thermodynamic quantities at the high $T$ limit are pushed beyond their corresponding SB limit and approach a new Tsallis limit. For a non-zero value of baryon chemical potential ($\mu_B = 600$ MeV), we can see from subplots (b), (d), (f), and (h) in Fig. \ref{thermo} that the thermodynamic quantities start to have non-zero values even at lower temperatures. Again, increasing $q$ from 1 to 1.10 results in a rise in the magnitude of these quantities, similar to the case of vanishing baryon chemical potential.
    \par
\begin{figure}
                \centering
                \includegraphics[]{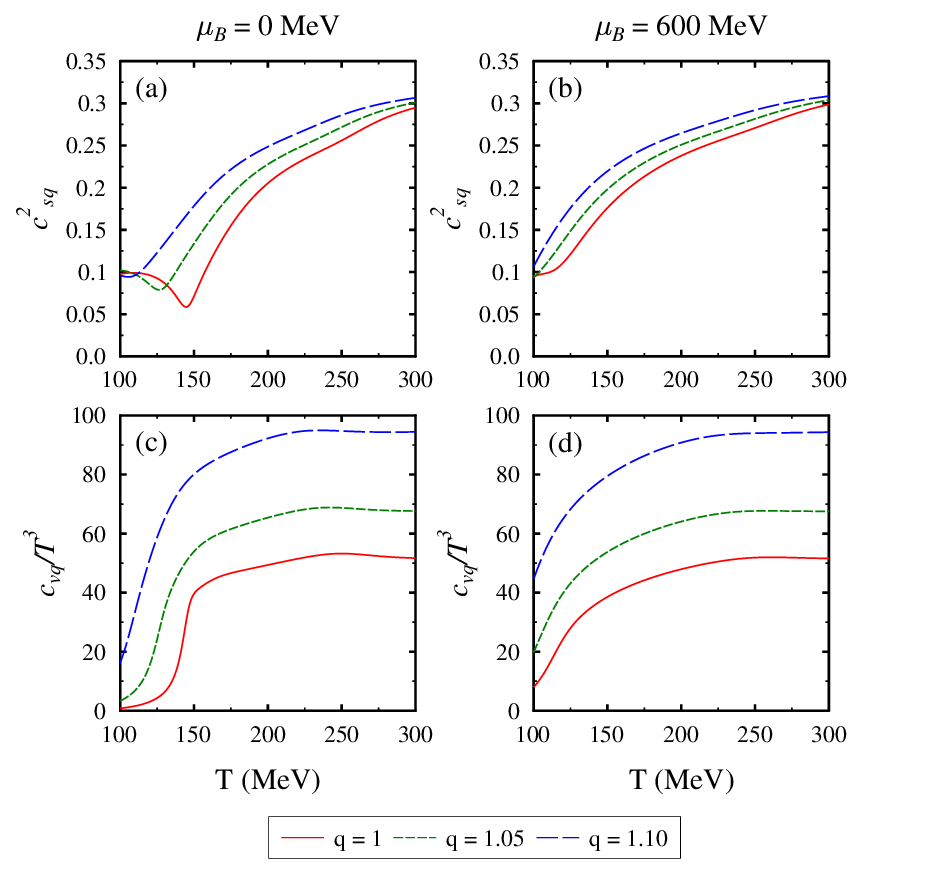}
                \caption{The speed of sound squared $c_{sq}^2$ and scaled specific heat $c_{vq}/T^3$ as a function of temperature $T$ for $q$ = 1, 1.05, and 1.10, at $\mu_B = 0$ MeV [in subplots (a) and (c)] and $\mu_B = 600$ MeV, $\mu_I = -30$ MeV, and $\mu_S=125$ MeV [in subplots (b) and (d)].}
                \label{sound}
            \end{figure}

 In Fig. \ref{sound}, we plot the scaled specific heat $c_{vq}/T^3$ and the speed of sound $c_{sq}^2$ as a function of temperature $T$ at baryon chemical potentials $\mu_B = 0$ and 600 MeV, for $q=1,1.05$, and 1.10. 
     For the case of vanishing $\mu_B$, we find that $c_{vq}/T^3$ exhibits a similar pattern to the other thermodynamic quantities. For $q=1$, it rises sharply near the transition temperature and reaches its SB limit at higher temperatures. For $q>1$, we observe that $c_{vq}/T^3$ surpasses its SB limit and reaches a higher Tsallis limit at high $T$. In subplot (d) of Fig. \ref{sound}, we observe that for non-zero $\mu_B$, the rise in $c_{vq}/T^3$ occurs at much lower temperatures before saturating to a constant value at higher temperatures. 
        As for the speed of sound, $c_{sq}^2$, we find that it reaches a minimum near the transition temperature and then increases to reach its SB limit at high $T$. We observe that the dip becomes less prominent, and its position is shifted towards lower temperatures as $q>1$. Additionally, for a fixed $T$, its value increases with increasing $q$. However, unlike the thermodynamic quantities discussed so far, $c_{sq}^2$ remains within the SB limit for all values of $q$ studied in the present work. This happens as $c_{sq}^2 = s_q/c_{vq}$, and both $s_q$ and $c_{vq}$ exhibit comparable increase at high temperature. This results in $c_{sq}^2$ always remaining under the SB limit. In addition, we find that the dip in $c_{sq}^2$ vanishes when the baryon chemical potential is increased to $\mu_B = 600$ MeV, as can be seen in Fig. \ref{sound}(d).
        \par
\begin{figure}
            	\centering
           \includegraphics[]{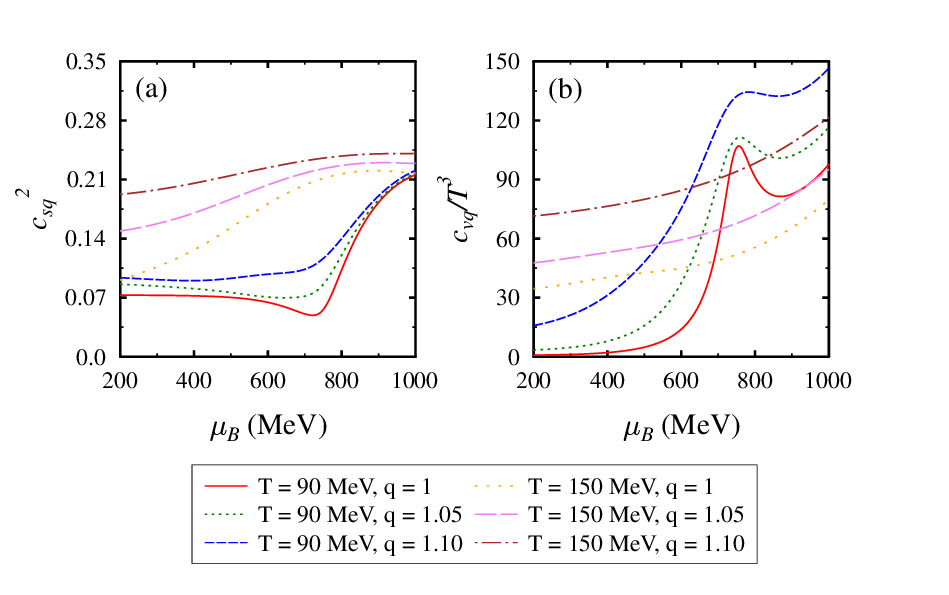}
            	\caption{\label{soundmu} The speed of sound squared $c_{sq}^2$ and scaled specific heat $c_{vq}/T^3$ as a function of baryon chemical potential $\mu_B$ for $q$ = 1, 1.05, and 1.10, at temperatures $T=90$ and 150 MeV.}
            \end{figure}

 In Figs. \ref{soundmu}(a) and (b), we have shown the dependence of the speed of sound squared $c_{sq}^2$ and scaled specific heat $c_{vq}/T^3$
  on the  baryon chemical potential $\mu_B$, for    
         $q = $ 1, 1.05, and 1.10 and  temperatures $T= 90$ and 150 MeV. The plots are shown for zero value of strangeness chemical potential $\mu_S$, isospin chemical potential $\mu_I$ and vector coupling constant $g_v$. For $q=1$ and $T = 90$ MeV, we observe that $c_{sq}^2$ starts from a low value at smaller $\mu_B$ and shows a dip before rising as $\mu_B$ increases. The dip in $c_{sq}^2$ may indicate the position of pseudo-critical temperature for chiral phase transition in the $T-\mu_B$ phase diagram \cite{heliu}. We find that as the value of $q$ is increased to 1.05 and 1.10, the dip shifts to lower values of $\mu_B$. This may signify a shift in the chiral transition to lower chemical potentials for nonextensive systems. An increase in the value of temperature 
        % to $T=150$ MeV,
    also shifts the dip in $c_{sq}^2$ to lower values of $\mu_B$. 
         %Again, increasing the nonextensivity in the system results in a higher value of $c_{sq}^2$.
          Similar kind of behaviour is also observed for the scaled specific heat $c_{vq}/T^3$ in Fig. \ref{soundmu}(b). For $T=90$ MeV, we find that $c_{vq}/T^3$ starts from a low value and rises with increasing baryon chemical potential before showing a peak at higher $\mu_B$. Increasing the value of $q$ shifts the position of the peak to lower $\mu_B$. At higher temperature of $T=150$ MeV, the peak in $c_v/T^3$ disappears.
        \par
\begin{figure}
                \centering
                \includegraphics[]{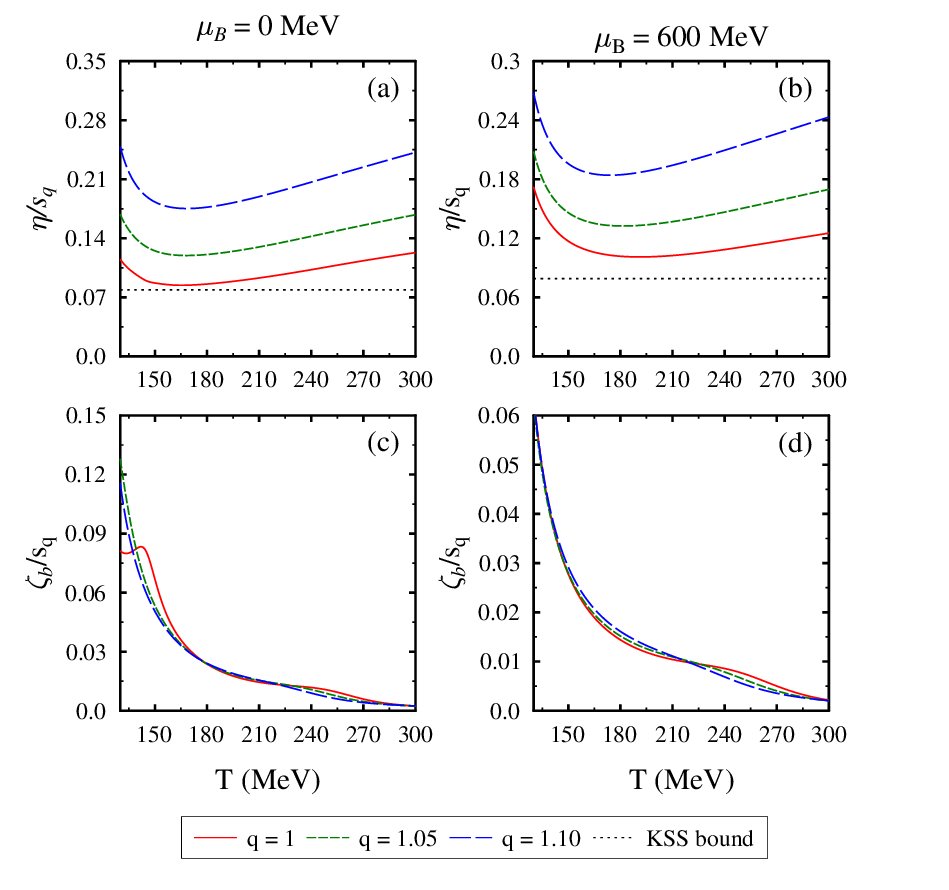}
                \caption{The specific shear viscosity $\eta/s_q$ and normalised bulk viscosity $\zeta_{b}/s_q$ as a function of temperature $T$ for $q$ = 1, 1.05, and 1.10, at $\mu_B = 0$ MeV [in subplots (a) and (c)] and $\mu_B = 600$ MeV, $\mu_I = -30$ MeV, and $\mu_S=125$ MeV [in subplots (b) and (d)].}
                \label{viscosity}
            \end{figure}

 Let us now begin the discussion of the viscous transport coefficients of the QCD matter. Shear viscosity $\eta$ and bulk viscosity $\zeta_b$ serve as important parameters to describe the hydrodynamical evolution of the QCD medium, thereby impacting phenomenological observables such as correlation functions and elliptical flow \cite{romat,olli}. In subplots (a) and (b) of Fig. \ref{viscosity}, we present the temperature dependence of the shear viscosity to entropy ratio $\eta/s_q$ in the $q-$PCQMF model for the value of $q = $ 1, 1.05, and 1.10. The results are shown for baryon chemical potentials $\mu_B = 0$ and 600 MeV. At vanishing baryon chemical potential, $\eta/s_q$ approaches the KSS bound ($1/4\pi$) and shows a minimum near the transition temperature and increases slowly thereafter. For temperatures below the transition temperature, $\eta/s_q$ grows and diverges as $T\rightarrow0$. This is similar to the findings observed in the PNJL model \cite{ghosh2015,saha2018}. Beyond the transition temperature, $\eta/s_q$ grows slowly and exhibits similarities to a fluid experiencing a phase change from liquid to gas \cite{schaefer2009}. As for the impact of the $q$ parameter on $\eta/s_q$, we find that its reaction to $q$ increases with temperature. The effective constituent quark mass decreases with temperature, leading to an increase in the value of $\eta/s_q$ at high temperatures. This decrease in the effective quark masses is more for systems with higher nonextensivity, i.e., $q>1$. Hence, resulting in an enhanced value of $\eta/s_q$ for $q>1$. This is similar to the observations in the $q-$PNJL model \cite{zhao2020} where $\eta$ is increased for values of $q>1$. In Ref. \cite{rath2024}, $\eta/s_q$ is found to have a little increase due to the nonextensive behaviour of the medium. Also, we find that the magnitude of the minima in $\eta/s_q$ increases for higher $q$ values. We observe the qualitative nature of $\eta/s_q$ remains the same, and the magnitude of the minima is increased for the value of baryon chemical potential $\mu_B = 600$ MeV. The effect of increasing $q$ remains the same as in the case of vanishing chemical potential.
        \par Fig. \ref{viscosity}(c) and (d) shows the bulk viscosity to entropy density ratio $\zeta_b/s_q$ as a function of temperature $T$ with baryon chemical potential fixed at $\mu_B = 0$ and 600 MeV, for $q=$ 1, 1.05, and 1.10. The bulk viscosity is an important parameter, especially due to its connection to the conformal symmetry of the system \cite{arnold06}. At low temperatures, $\zeta_b/s_q$ is large due to the comparative magnitudes of $\zeta_b$ and $s_q$. For $\mu_B = 0$ MeV, we note that $\zeta_b/s_q$ shows a slight peak close to the transition temperature and then slowly decreases to zero at higher temperatures. This vanishing value of $\zeta_b/s_q$ can be explained by the significant increase in $s_q$ relative to $\zeta_b$, suggesting that the system achieves conformal symmetry at high temperatures. The PNJL model \cite{saha2017} and the PLSM model \cite{tawfik} also show a similar trend for $\zeta_b/s_q$. Increasing the nonextensivity ($q>1$), we find that the peak in $\zeta_b/s_q$ disappears as it gradually goes to zero at high $T$. Additionally, unlike $\eta/s_q$, we find that the normalised bulk viscosity $\zeta_b/s_q$ decreases as the nonextensivity of the medium is increased. This is similar to the findings in Ref. \cite{rath2024}. The decreasing value of $\zeta_b/s_q$ indicates that the system gets closer to the conformal limit as the degree of nonextensivity of the medium is increased. Furthermore, at finite baryon chemical potential of $\mu_B=600$ MeV, we observe that $\zeta_b/s_q$ shows no peak, and its magnitude gets smaller as can be seen in subplot (d) of Fig. \ref{viscosity}. In addition, we observe that increasing the value of $q$ results in a slight increase in $\zeta_b/s_q$ at lower temperatures, while a slight decrease at higher temperatures.
        \par
\begin{figure}
                \centering
                \includegraphics[]{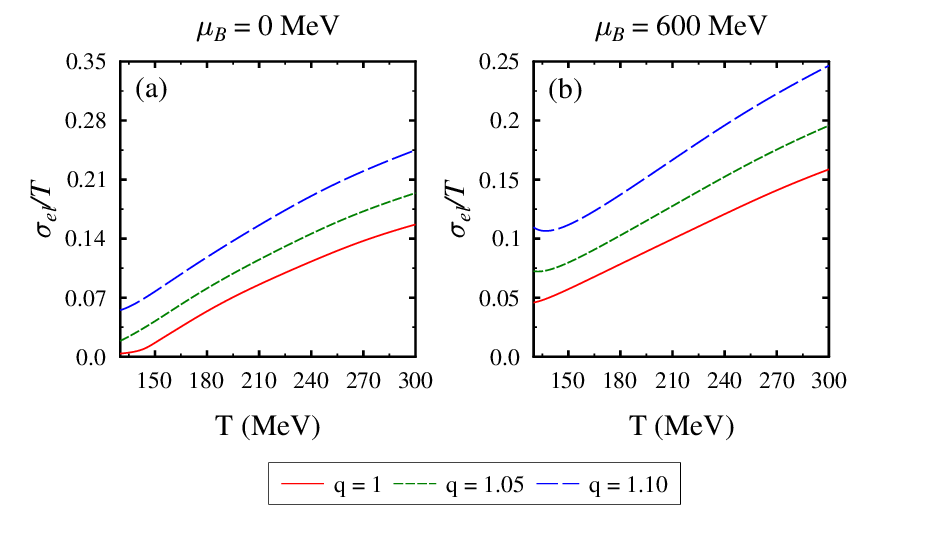}
                \caption{The normalised electrical conductivity $\sigma_{el}/T$ as a function of temperature $T$ for $q$ = 1, 1.05, and 1.10, at $\mu_B = 0$ MeV [in subplot (a)] and $\mu_B = 600$ MeV, $\mu_I = -30$ MeV, and $\mu_S=125$ MeV [in subplot (b)].}
                \label{sigmael}
            \end{figure}

 Now, we come to the discussion of the conductive transport coefficients, namely, the electrical conductivity $\sigma_{el}$ and the thermal conductivity $\kappa$ in the $q-$PCQMF model. In Fig, \ref{sigmael}, we show temperature variation of $\sigma_{el}/T$ for both zero as well as the finite value of baryon chemical potential at $q=$ 1, 1.05, and 1.10. The electrical conductivity $\sigma_{el}$ serves as a vital tool to understand the electromagnetic interactions in QCD matter. For $\mu_B=0$ MeV, we find $\sigma_{el}$ to increase with temperature $T$. This can be attributed to the deconfinement of quarks at high $T$, allowing them to move around easily and enhance electrical conductivity. The NJL model \cite{ghosh2019} and PNJL model \cite{saha2018} report similar findings for $\sigma_{el}/T$. Regarding the impact of nonextensivity $q$ on $\sigma_{el}/T$, we observe that increasing $q$ results in increased electrical conductivity, even at lower temperatures. The electrical conductivity, $\sigma_{el}$, is found to increase with the introduction of nonextensivity in the kinetic theory approach \cite{rath2023} and in the $q-$PNJL model \cite{zhao2020}. This may be due to a decrease in the effective quark mass for the system with higher nonextensivity ($q>1$). Another reason could be the tendency of the quarks to become deconfined at lower temperatures for nonextensive systems, as was pointed out in Fig. \ref{phi}. Increasing the baryon chemical potential $\mu_B$ to 600 MeV results in an increased magnitude of $\sigma_{el}/T$ at lower temperatures. We find $\sigma_{el}/T$ to increase with the rise in the degree of nonextensivity ($q>1$), similar to the case of $\mu_B = 0$ MeV.
        \par
\begin{figure}
                \centering
                \includegraphics[]{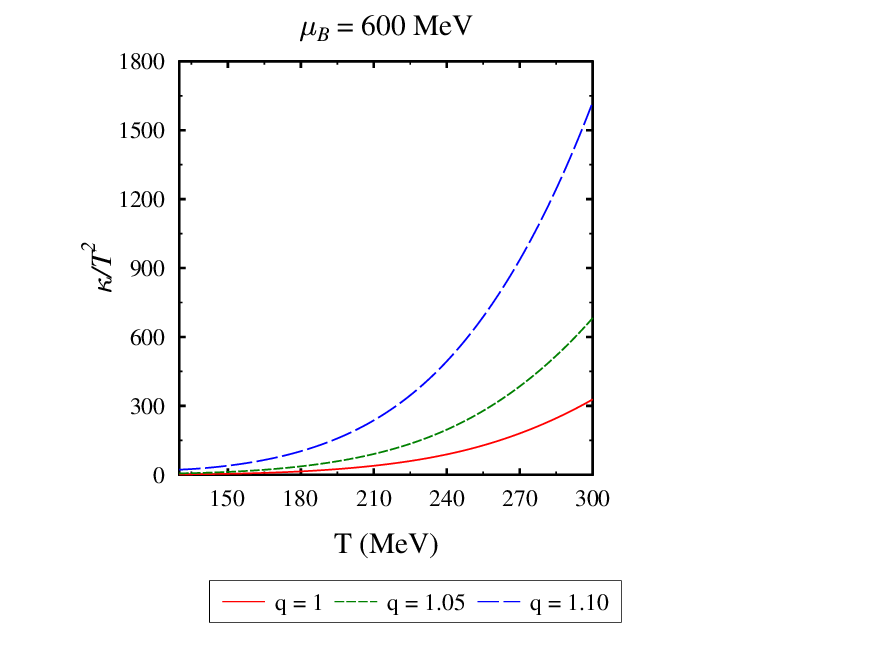}
                \caption{The variation of normalised thermal conductivity $\kappa/T^2$ with temperature $T$ for $q$ = 1, 1.05, and 1.10, at $\mu_B = 600$ MeV, $\mu_I = -30$ MeV, and $\mu_S=125$ MeV}
                \label{kappaq}
            \end{figure}

 We have also computed the thermal conductivity $\kappa/T^2$, which is associated with the heat flow of the QGP. In Fig. \ref{kappaq}, we plot $\kappa/T^2$ as a function of $T$ for $\mu_B = 600$ MeV, $\mu_I = -30$ MeV, $\mu_S=125$ MeV, at $q = 1$, 1.05, and 1.10. According to Eq. (\ref{kappa}), $\kappa/T^2$ diverges at $\mu_B=0$ MeV due to the diverging nature of the heat function $h$ at vanishing chemical potential. As the temperature of the system increases, the value of $h$ increases, leading to an efficient transmission of heat within the QGP, thereby enhancing thermal conductivity. As for the impact of the $q$ parameter, we find that similar to other transport coefficients, $\kappa/T^2$ increases for $q>1$ with the influence more noticeable in the high $T$ region. Again, this is similar to the observations in Ref. \cite{rath2023}. 
        \par
\begin{figure}
            	\centering
      \includegraphics[]{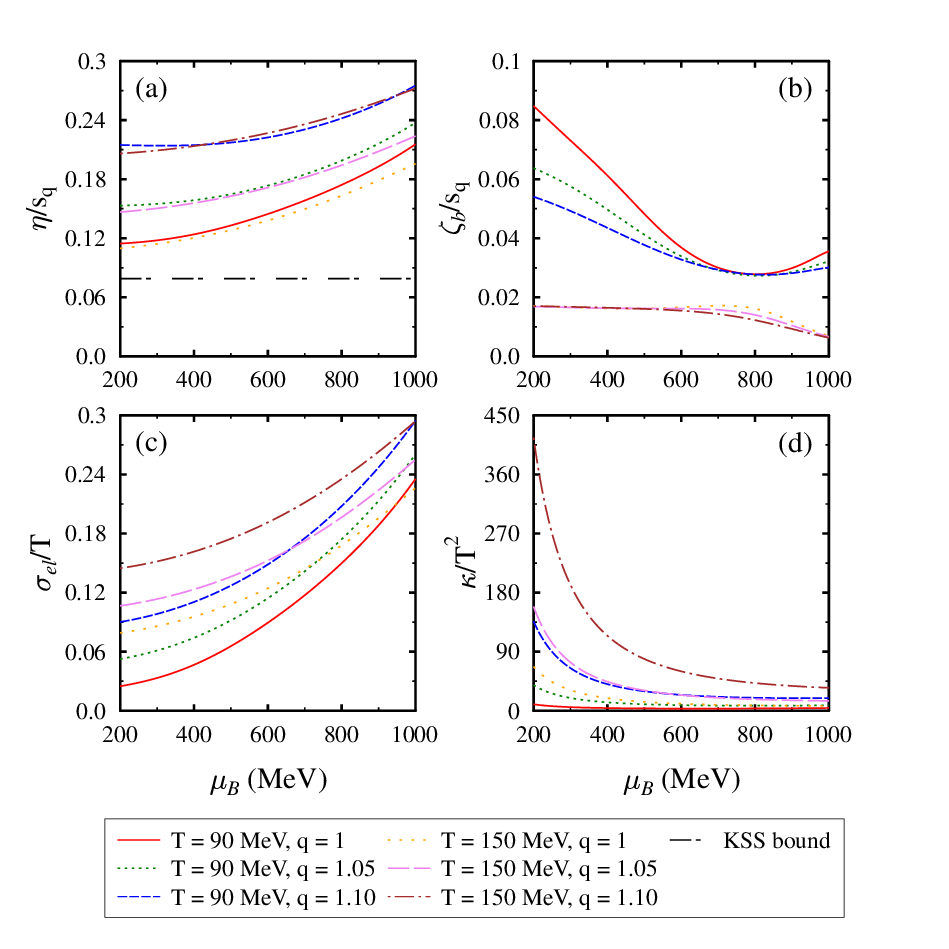}
            	\caption{\label{transmu} The specific shear viscosity $\eta/s_q$, normalised bulk viscosity $\zeta_{b}/s_q$, normalised electrical conductivity $\sigma_{el}/T$, and normalised thermal conductivity $\kappa/T^2$ as a function of baryon chemical potential $\mu_B$ for $q$ = 1, 1.05, and 1.10, at temperature $T=90$  and 150 MeV.}
            \end{figure}

 Finally, we have shown the baryon chemical potential dependence of the transport coefficients in Fig. \ref{transmu} for different values of $q$ at temperatures $T=150$  and 200 MeV and $\mu_S = \mu_I = g_v =0$. In Fig. \ref{transmu}(a), we find that $\eta/s_q$ increases with increasing value of $\mu_B$. The value of $\eta/s_q$ is found to increase with increasing nonextensivity in the system (higher $q$ values). A contrasting behaviour is observed in $\zeta_b/s_q$, which is found to decrease with increasing $\mu_B$ and $q$, as shown in Fig. \ref{transmu}(b). Also, we find that increasing the temperature results in lower values of $\zeta_b/s_q$, suggesting conformal symmetry restoration. As for the electrical conductivity $\sigma_{el}/T$ in Fig. \ref{transmu}(c), we find that it starts from a low value and rises with $\mu_B$. While the thermal conductivity $\kappa/T^2$ is observed to start from a higher value at small $\mu_B$ and decreases with increasing $\mu_B$ as shown in Fig. \ref{transmu}(d). The high value of $\kappa/T^2$ at small $\mu_B$ is due to its diverging nature at $\mu_B=0$ as pointed out earlier. As for the impact of nonextensivity, we observe that both electrical and thermal conductivity increases for higher $q$ values and the influence is more prominent at lower $\mu_B$.
	\section{\label{summary}Summary and Conclusion}
        In this study, we have used the nonextensive Polyakov chiral SU(3) quark mean field ($q-$PCQMF) model for three quark flavours ($u,d,s$) to calculate the transport coefficients such as the shear $\eta$ and bulk viscosity $\zeta_b$ as well as the electrical $\sigma_{el}$ and thermal conductivity $\kappa$ as a function of temperature $T$. We have evaluated the transport coefficients for both zero and finite baryon chemical potential ($\mu_B = 600$ MeV), taking into account the finite values of isospin ($\mu_I=-30$ MeV) and strangeness chemical potentials ($\mu_S = 125$ MeV). The value of the nonextensivity parameter $q$ used are 1, 1.05., and 1.10. In order to examine the transport coefficients while taking into account the nonextensivity, we have computed the $q$-dependent thermodynamic quantities such as pressure density $p_q/T^4$, energy density $\epsilon_q/T^4$, entropy density $s_q/T^3$, trace anomaly $(\epsilon_q-3p_q)/T^4$, speed of sound squared $c_{sq}^2$, and specific heat $c_{vq}/T^3$. We find that at higher temperatures, all thermodynamic quantities except $c_{sq}^2$ approach the $q-$dependent Tsallis limit instead of their usual SB limit. Due to an unexpected cancellation, the high-temperature limit of $c_{sq}^2$ remains unaffected. We also found that the effective quark masses decrease as the nonextensivity of the system increases, possibly due to the shift of chiral symmetry restoration to lower temperatures. For vanishing baryon chemical potential, the temperature variation of the ratio of shear viscosity to entropy density $\eta/s_q$ shows that it approaches the KSS bound and a minimum near the transition temperature, rising thereafter. On the other hand,  the bulk viscosity to entropy ratio $\zeta_{b}/s_q$ exhibits a small peak near the transition temperature and vanishes at higher temperatures. Furthermore, we find that both the electrical conductivity $\sigma_{el}$ and the thermal $\kappa$ conductivity increase monotonically with temperature. Regarding the impact of the nonextensive $q$ parameter, we find that the transport coefficients $\eta/s_q$, $\sigma_{el}/T$, and $\kappa/T^2$ get enhanced while $\zeta_b/s_q$ get diminished for systems with $q>1$. The impact of nonextensivity on the transport coefficients is found to be more prominent in the high $T$ range. In addition, we have also studied all the transport coefficients at finite values of chemical potentials. We found that increasing the chemical potentials leads to a larger magnitude of the transport coefficients at lower temperatures, possibly owing to the early restoration of chiral symmetry for systems with finite density. In future, we aim to utilise the $q-$PCQMF model to study the transport coefficient in a magnetised nonextensive QCD medium.

\section{ACKNOWLEDGMENT}
The authors sincerely acknowledge the support toward this work from the Ministry of Science and Human Resources (MHRD), Government of India, via an Institute fellowship under the Dr B R Ambedkar National Institute of Technology Jalandhar.

	\section*{Appendix}

The coupled equations  are obtained by minimizing the thermodynamic potential density, $\Omega_q$, with respect to the various fields of the $q$-PCQMF model and are given as

\begin{eqnarray}\label{sigma1}
	&&\frac{\partial \Omega_q}{\partial \sigma}= k_{0}\chi^{2}\sigma-4k_{1}\left( \sigma^{2}+\zeta^{2}
	+\delta^{2}\right)\sigma-2k_{2}\left( \sigma^{3}+3\sigma\delta^{2}\right)
	-2k_{3}\chi\sigma\zeta \nonumber\\
	&-&\frac{d}{3} \chi^{4} \bigg (\frac{2\sigma}{\sigma^{2}-\delta^{2}}\bigg )
	+\left( \frac{\chi}{\chi_{0}}\right) ^{2}m_{\pi}^{2}f_{\pi}- 
	\left(\frac{\chi}{\chi_0}\right)^2m_\omega\omega^2
	\frac{\partial m_\omega}{\partial\sigma}\nonumber\\
	&-&\left(\frac{\chi}{\chi_0}\right)^2m_\rho\rho^2 
	\frac{\partial m_\rho}{\partial\sigma}
	-\sum_{i=u,d} g_{\sigma}^i\rho_{q,i}^{s} = 0 ,
\end{eqnarray}
\begin{eqnarray}
	&&\frac{\partial \Omega_q}{\partial \zeta}= k_{0}\chi^{2}\zeta-4k_{1}\left( \sigma^{2}+\zeta^{2}+\delta^{2}\right)
	\zeta-4k_{2}\zeta^{3}-k_{3}\chi\left( \sigma^{2}-\delta^{2}\right)-\frac{d}{3}\frac{\chi^{4}}{{\zeta}}\nonumber\\
	&+&\left(\frac{\chi}{\chi_{0}} \right)
	^{2}\left[ \sqrt{2}m_{K}^{2}f_{K}-\frac{1}{\sqrt{2}} m_{\pi}^{2}f_{\pi}\right]-\left(\frac{\chi}{\chi_0}\right)^2m_\phi\phi^2 
	\frac{\partial m_\phi}{\partial\zeta}
	-\sum_{i=s} g_{\zeta}^i\rho_{q,i}^{s} \nonumber\\&=& 0 ,
	\label{zeta}
\end{eqnarray}
\begin{eqnarray}
	&&\frac{\partial \Omega_q}{\partial \delta}=k_{0}\chi^{2}\delta-4k_{1}\left( \sigma^{2}+\zeta^{2}+\delta^{2}\right)
	\delta-2k_{2}\left( \delta^{3}+3\sigma^{2}\delta\right) +\mathrm{2k_{3}\chi\delta
		\zeta} \nonumber\\
	& + &  \frac{2}{3} d \chi^4 \left( \frac{\delta}{\sigma^{2}-\delta^{2}}\right)
	-\sum_{i=u,d} g_{\delta}^i\rho_{q,i}^{s} = 0 ,
	\label{delta}
\end{eqnarray}
\begin{eqnarray}
	&&\frac{\partial \Omega_q}{\partial \chi}=\mathrm{k_{0}\chi} \left( \sigma^{2}+\zeta^{2}+\delta^{2}\right)-k_{3}
	\left( \sigma^{2}-\delta^{2}\right)\zeta + \chi^{3}\left[1
	+{\rm {ln}}\left( \frac{\chi^{4}}{\chi_{0}^{4}}\right)  \right]
	+(4k_{4}-d)\chi^{3}
	\nonumber\\
	&-&\frac{4}{3} d \chi^{3} {\rm {ln}} \Bigg ( \bigg (\frac{\left( \sigma^{2}
		-\delta^{2}\right) \zeta}{\sigma_{0}^{2}\zeta_{0}} \bigg )
	\bigg (\frac{\chi}{\mathrm{\chi_0}}\bigg)^3 \Bigg )+
	\frac{2\chi}{\chi_{0}^{2}}\left[ m_{\pi}^{2}
	f_{\pi}\sigma +\left(\sqrt{2}m_{K}^{2}f_{K}-\frac{1}{\sqrt{2}}
	m_{\pi}^{2}f_{\pi} \right) \zeta\right] \nonumber\\
	&-& \frac{\chi}{{\chi^2_0}}({m_{\omega}}^2 \omega^2+{m_{\rho}}^2\rho^2)  = 0,
	\label{chi}
\end{eqnarray}
\begin{eqnarray}
	\frac{\partial \Omega_q}{\partial \omega}=\frac{\chi^2}{\chi_0^2}m_\omega^2\omega+4g_4\omega^3+12g_4\omega\rho^2
	&-&\sum_{i=u,d}g_\omega^i\rho_{q,i}=0,
	\label{omega} 
\end{eqnarray}
\begin{eqnarray}
	\frac{\partial \Omega_q}{\partial \rho}=\frac{\chi^2}{\chi_0^2}m_\rho^2\rho+4g_4\rho^3+12g_4\omega^2\rho&-&
	\sum_{i=u,d}g_\rho^i\rho_{q,i}=0, 
	\label{rho} 
\end{eqnarray}
\begin{eqnarray}
	\frac{\partial \Omega_q}{\partial \phi}=\frac{\chi^2}{\chi_0^2}m_\phi^2\phi+8g_4\phi^3&-&
	\sum_{i=s}g_\phi^i\rho_{q,i}=0,
	\label{phi}  
\end{eqnarray}
\begin{eqnarray}
	\hspace*{0.4cm} 
	%\vspace*{-0.4cm} 
	\frac{\partial \Omega_q}{\partial \Phi} =\bigg[\frac{-a(T)\bar{\Phi}}{2}-\frac{6b(T)
		(\bar{\Phi}-2{\Phi}^2+{\bar{\Phi}}^2\Phi)
	}{1-6\bar{\Phi}\Phi+4(\bar{\Phi}^3+\Phi^3)-3(\bar{\Phi}\Phi)^2}\bigg]T^4
	-\sum_{i=u,d,s}\frac{2k_BTN_C}{(2\pi)^3}
	\nonumber\\
	\int_0^\infty d^3k 
	\bigg[\frac{\exp_q\left(\frac{-(E_i^*(k)-{\mu_i}^{*})}{k_BT}\right)}{\left(1+\exp_q\left(\frac{-3(E_i^*(k)-{\mu_i}^{*})}{k_BT}\right)+3\Phi \exp_q\left(\frac{-(E_i^*(k)-{\mu_i}^{*})}{k_BT}\right)
		+3\bar{\Phi}\exp_q\left(\frac{-2(E_i^*(k)-{\mu_i}^{*})}{k_BT}\right)\right)^q}
	\nonumber\\
	+\frac{\exp_q\left(\frac{-2(E_i^*(k)+{\mu_i}^{*})}{k_BT}\right)}{\left(1+\exp_q\left(\frac{-3(E_i^*(k)+{\mu_i}^{*})}{k_BT}\right)
		+3\bar{\Phi} \exp_q\left(\frac{-(E_i^*(k)+{\mu_i}^{*})}{k_BT}\right)+3\Phi \exp_q\left(\frac{-2(E_i^*(k)+{\mu_i}^{*})}{k_BT}\right)\right)^q}\bigg] \nonumber \\=0,
	\label{Polyakov} 
\end{eqnarray}
and
\begin{eqnarray}
	\frac{\partial \Omega_q}{\partial \bar{\Phi}} =\bigg[\frac{-a(T)\Phi}{2}-\frac{6b(T)
		(\Phi-2{\bar{\Phi}}^2+{\Phi}^2\bar{\Phi})
	}{\mathrm{1-6\bar{\Phi}\Phi+4(\bar{\Phi}^3+\Phi^3)-3(\bar{\Phi}\Phi)^2}}\bigg]T^4
	-\sum_{i=u,d,s}\frac{2k_BTN_C}{(2\pi)^3}
	\nonumber\\
	\int_0^\infty d^3k 
	\bigg[\frac{\exp_q\left(\frac{-2(E_i^*(k)-{\mu_i}^{*})}{k_BT}\right)}{\left(1+\exp_q\left(\frac{-3(E_i^*(k)-{\mu_i}^{*})}{k_BT}\right)+3\Phi \exp_q\left(\frac{-(E_i^*(k)-{\mu_i}^{*})}{k_BT}\right)
		+3\bar{\Phi}\exp_q\left(\frac{-2(E_i^*(k)-{\mu_i}^{*})}{k_BT}\right)\right)^q}
	\nonumber\\
	+\frac{\exp_q\left(\frac{-(E_i^*(k)+{\mu_i}^{*})}{k_BT}\right)}{\left(1+\exp_q\left(\frac{-3(E_i^*(k)+{\mu_i}^{*})}{k_BT}\right)
		+3\bar{\Phi} \exp_q\left(\frac{-(E_i^*(k)+{\mu_i}^{*})}{k_BT}\right)+3\Phi \exp_q\left(\frac{-2(E_i^*(k)+{\mu_i}^{*})}{k_BT}\right)\right)^q}\bigg] \nonumber \\=0. 
	\label{Polyakov conjugate} 
\end{eqnarray}

            \begin{figure}
                \centering
                \includegraphics[]{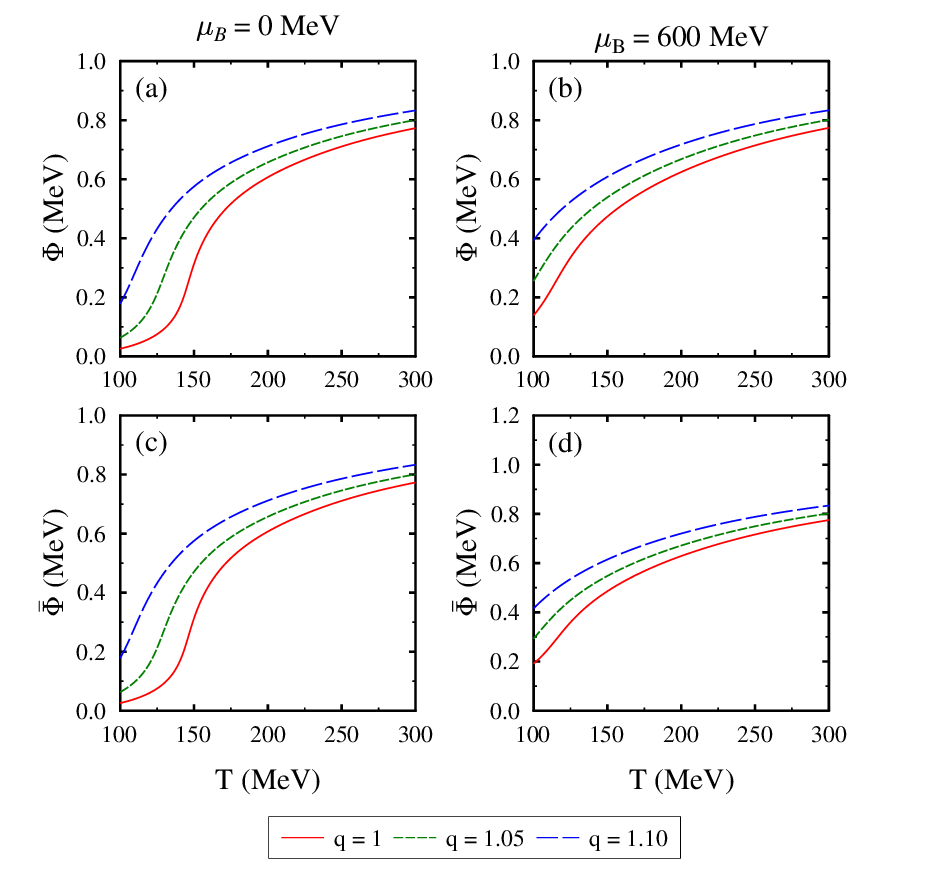}
                \caption{The Polyakov fields $\Phi$ and $\bar{\Phi}$ are plotted as a function of temperature $T$ for the value of nonextensivity parameter $q$ = 1, 1.05, and 1.10, at baryon chemical potential $\mu_B = 0$ MeV [in subplots (a) and (c)], and baryon chemical potential $\mu_B = 600$ MeV, isospin chemical potential $\mu_I = -30$ MeV, and strangeness chemical potential $\mu_S=125$ MeV [in subplots (b) and (d)].}
                \label{phi}
            \end{figure}

\end{document}